\documentclass[a4paper,fleqn,usenatbib]{mnras}

\usepackage[T1]{fontenc}
\usepackage{ae,aecompl}

\usepackage{graphicx}	
\usepackage{amsmath}	
\usepackage{amssymb}	
\usepackage{array}	
\usepackage{hyperref}

\newcommand{\nugc}{{\scriptsize$\nu^{2}$GC}}
\newcommand{\Soltan}{So\l tan's~}

\defcitealias{nu2gc}{M16}
\defcitealias{KH00}{KH00}
\defcitealias{KS13}{KS13}
\graphicspath{{figures/}}

\title[Theoretical predictions of the Eddington ratio]{Slowing down of cosmic growth of supermassive black holes: Theoretical prediction of the  Eddington ratio distribution}

\author[H.~Shirakata et al.]
{Hikari~Shirakata,$^{1,2}$\thanks{E-mail: shirakata@astro1.sci.hokudai.ac.jp} Toshihiro~Kawaguchi,$^3$\thanks{E-mail: kawaguchi@onomichi-u.ac.jp} Taira~Oogi,$^4$
    \newauthor Takashi~Okamoto,$^{1}$ and Masahiro~Nagashima $^5$ \\
  $^1$Department of Cosmosciences, Graduate School of Science, Hokkaido University, N10 W8, Kitaku, Sapporo, 060-0810, Japan\\
	$^2$The Technical Research Center, Tadano Ltd., 2217-13, Hayashi-machi, Takamatsu, Kagawa, 761-0301, Japan\\
  $^3$Department of Economics, Management and Information Science, Onomichi City University, 1600-2, Hisayamada, Onomichi, Hiroshima, 722-8506, Japan\\
  $^4$Kavli Institute for the Physics and Mathematics of the Universe (WPI), Todai Institutes for Advanced Study, the University of Tokyo, 5-1-5, Kashiwanoha, Kashiwa, 277-8583 Japan\\
  $^5$Faculty of Education, Bunkyo University, 3337, Minami-ogishima, Koshigaya, Saitama 343-8511, Japan\\
  }

\date{Accepted XXX. Received YYY; in original form ZZZ}

\pubyear{2018}

\begin{document}
  \label{firstpage}
  \pagerange{\pageref{firstpage}--\pageref{lastpage}}
  \maketitle

  \begin{abstract}
    We show the Eddington ratio distributions of
    supermassive black holes at a wide redshift range $(0 < z < 8)$
    obtained with a semi-analytic model of galaxy formation.
		The distribution is broadly consistent with observational estimates at low redshift.
		We find that the growth rate of black holes at higher redshift is more likely to exceed the Eddington limit
    because the typical gas fraction of the host galaxies is higher at higher redshift.
		We also find that the super-Eddington growth is more common for
		less massive supermassive black holes, supporting an idea
		that supermassive black holes have been formed via super-Eddington accretion.
		These results indicate the ``slowing down'' of cosmic growth of supermassive black holes:
		the growth of supermassive black holes with a higher Eddington ratio peaks at higher redshift.
		We also show the effect of the sample selection on the shape of the Eddington ratio distribution functions
		and find that shallower observations will miss active galactic nuclei with not only the smaller but also higher Eddington ratios.
   \end{abstract}




  \section{Introduction}
	\label{sec:intro}
	Previous studies suggest that almost all galaxies at $z \sim 0$
	have a supermassive black hole (SMBH) at their centre \citep[e.g.][]{Magorrian98}.
	It is believed that SMBHs grow mainly via the gas accretion \citep{Salpeter64,Lynden-Bell69}.
  \footnote{SMBHs would also grow by the BH-BH coalescence \citep[e.g.][]{KH00}.
	However, the contribution would be negligible for the cosmic growth of SMBHs,
	based on the statistical spin estimation for the observed local SMBHs \citep{Marconi04} and
	semi-analytic approach \citep[at least $z > 1$;][]{Enoki04}.}
	The growth can be observed as an active galactic nucleus (AGN),
	which emits light when the material gets accreted onto the SMBH.
	AGNs also have the kinetic jet and/or outflow, which possibly affect the gas kinematics surrounding the SMBHs
	and the star formation activity in AGN host galaxies.

	As an indicator for how rapid an SMBH grows,
	the accretion rate and luminosity, which are normalised by the Eddington limit, has been employed.
	The Eddington luminosity and Eddington accretion rate are defined as
	\begin{align}
    &L_\mathrm{Edd} = \frac{4\pi c G m_p}{\sigma_\mathrm{T}} M_\mathrm{BH}, \\
    &\dot{M}_\mathrm{Edd} = L_\mathrm{Edd} / c^2,
  \end{align}
	where $G, m_p, \sigma_\mathrm{T}$, $c$, and $M_\mathrm{BH}$ are the gravitational constant, proton mass,
	cross-section for the Thomson scattering, the speed of light, and the mass of a BH, respectively.
	The gravitational force balances radiative pressure on the accreted gas
	in a spherical accretion and illumination case,
	when the accretion rate, $\dot{M}_\mathrm{BH}$, is $\sim 10 \dot{M}_\mathrm{Edd}$
	(i.e. $L_\mathrm{bol} \sim L_\mathrm{Edd}$, assuming that
	the radiation efficiency of sub-Eddington accretion rate is about $0.1$).

	\cite{Soltan82} finds that the mass density of local SMBHs
	is explained by integrating luminosity functions (LFs) of QSOs
	(i.e. the brightest class of AGNs) with the radiation efficiency,
	$\epsilon \equiv L_\mathrm{bol} / \dot{M}_\mathrm{BH} c^2 \sim 0.1$
	($L_\mathrm{bol}$ is the bolometric luminosity of AGNs).
	This argument also includes the important suggestion for the SMBH growth:
	the ``super-Eddington growth'' is not required.
	The accretion rate normalised by the Eddington accretion rate, $\dot{m}$, affects the
	properties of the accretion discs and determines the value of $\epsilon$
	because the structure and temperature distribution of the accretion discs depend on $\dot{m}$.
	The value of $\epsilon \sim 0.1$ is for the standard accretion disc \citep{SS73} with a sub-Eddington accretion rate,
	while a super-Eddington accretion ($\dot{M}_\mathrm{BH}/\dot{M}_\mathrm{Edd} \equiv \dot{m} \gg 10$) results in a much smaller $\epsilon$
	\citep{Abramowicz88}.
	The \Soltan argument and updated results later on \citep[e.g.][]{YT02}, thus,
	suggest that local SMBHs have grown mainly via sub-Eddington accretions.

	However, based on observed accretion rates and inferred durations
	of super-Eddington AGNs together with the observed trend
	of higher Eddington ratios at higher redshift \citep{MD04},
	\cite{Kawaguchi04June} argue that SMBHs have grown via
	super-Eddington accretion in the early Universe.
	The number density of QSOs at the ``knee'' of the QSO LFs
	mostly governs the integrated energy density.
	Namely, little information at other luminosity ranges
	(such as lower luminosity than the ``knee'')
	is included in the integrated values,
	resulting in little constraints on less massive SMBHs.
	The luminosity at the knee corresponds to about $10^{8-9} M_\odot$,
	assuming that all AGNs radiate at the Eddington luminosity.
	Therefore, the \Soltan argument seems not to reject the super-Eddington growth of SMBHs
	with $M_\mathrm{BH} \lesssim 10^8 M_\odot$
	and the contribution of the super-Eddington accretion for the cosmic growth of less massive SMBHs,
	which are the majority of SMBHs, is unclear.

	Estimations of the Eddington ratio distribution functions (ERDFs)
	based on observational data
	suffer severely from a flux-limit, a matter of completeness,
	and a priori assumptions (e.g. log-normal distribution of ERDFs).
	For instance, ERDFs at $z = 2.15$ estimated by \citet[hereafter KS13]{KS13}
	have a $\pm 1 \sigma$ uncertainty in the number density
  for a given $\lambda_\mathrm{Edd} \equiv L_\mathrm{bol}/L_\mathrm{Edd}$ of more than 2 dex.
	Despite the difficulties for obtaining the Eddington ratio,
	observations reveal the evolution of the ERDFs
	and relations between Eddington ratio and properties of the AGNs and their host galaxies.
	\citetalias{KS13} estimate the ``growth timescale'' defined as
	$t_s \ln (M_\mathrm{BH}/M_\mathrm{seed}) / [10 \epsilon E(L_\mathrm{bol}/L_\mathrm{Edd} \mid M_\mathrm{BH}, z)]$,
	where $t_s, M_\mathrm{seed}$, and $ E(L_\mathrm{bol}/L_\mathrm{Edd} \mid M_\mathrm{BH}, z)$
	are the Salpeter time, the seed BH mass and the mean value of $L_\mathrm{bol}/L_\mathrm{Edd}$ at a fixed $M_\mathrm{BH}$ and $z$,
	respectively, and $\epsilon = 0.1$ and $M_\mathrm{seed} = 10^6 M_\odot$.
	They find that the timescale for type-1 QSOs with $M_\mathrm{BH} > 5 \times 10^8 M_\odot$ at $z > 1.8$,
	is comparable to or longer than the age of the Universe,
	which suggests that such an SMBH population
	would have grown with the higher Eddington ratio at higher redshift.
	Indeed, recent observations find QSOs at $z > 6$ with $M_\mathrm{BH} > 10^8 M_\odot$
	which are growing at $\lambda_\mathrm{Edd} \gtrsim 1$ \citep{Mortlock11, Wu15, Banados18}.
	Other observations \citep[e.g.][]{Nobuta12, Shen12, Lusso12, Bongiorno16, Aird18} also show
	the Eddington ratio becomes higher at higher redshift.
	Theoretical studies for the formation of galaxies and SMBHs
	(i.e. cosmological simulations and semi-analytic models of galaxy formation; hereafter SA models)
	also investigate the evolution of the ERDFs and find that super-Eddington growth is more
	common for SMBHs at higher redshift \citep[e.g.][]{Hirschmann14, Sijacki15}.

	The contribution of the super-Eddington accretion to the cosmic growth of SMBHs is important
	to discuss the mass of seed black holes.
	Observations have found luminous quasars at $z > 6$, of which SMBH masses estimated as $> 10^{9} M_\odot$
  \citep[e.g.][]{Mortlock11,Wu15,Banados18}.
	There are three possibilities for explaining the existence of such massive SMBHs in the early Universe:
	(1) their seed BH masses are large, namely, $> 10^5 M_\odot$,
	(2) their seed BHs are small ($10^{1-3} M_\odot$) and formed soon after the birth of the Universe,
	and then grow continuously with the Eddington-limited accretion, and
	(3) their seed BHs are small ($10^{1-3} M_\odot$) and grow via super-Eddington accretion.
  However, ``heavy seed BHs'' should form only in special conditions \citep[e.g.][]{OSH08}
	and thus such ``heavy seed BHs'' should not be dominant \citep[see also][]{Shirakata16}.
	Although the formation time of the seed BH of SMBHs with $> 10^9 M_\odot$ at $z > 6$ is unclear,
	the super-Eddington growth might play a key role in the early growth of SMBHs
  since the typical Eddington ratio becomes higher at higher redshift \citep{Wu15}.

	In this paper, we present a theoretical prediction for the ERDFs obtained with an SA model,
	``\textit{New Numerical Galaxy Catalogue ($\nu^2 GC$)}''
	that has explained many observational properties of galaxies and AGNs,
	such as the evolution of LFs and stellar mass functions, local $M_\mathrm{BH}$ -- $M_\mathrm{bulge}$ relation,
	and size/velocity -- magnitude relations of local galaxies.
	We focus on the importance of the super-Eddington growth for different SMBH mass ranges.
	We also show the effect of the sample selection for obtaining the ERDFs
	and the dependence of the ERDFs on AGN host galaxies,
	which has not been studied in depth.
	The modelling and values of adjustable parameters are the same as \cite{Shirakata19},
	which present the detailed model description.
	This paper is organised as follows.
	In Sec. \ref{sec:model}, we briefly describe the growth model of SMBHs.
	In Sec. \ref{sec:results}, we show the ERDF obtained with the fiducial model \citep{Shirakata19}
	and show the effect of the sample selections.
	In Sec. \ref{sec:discussion},
	we discuss the properties of host galaxies and their relations to the Eddington ratio.
	Finally, we summarise the results in Sec. \ref{sec:conclusion}.
	Unless otherwise stated, we employ the $N$-body simulation \citep{Ishiyama15} with the largest box size $1120 h^{-1}$ Mpc
	and $8192^3$ particles (the smallest halo mass is $8.79 \times 10^9 M_\odot$).

  \section{SMBH Growth Model}
  \label{sec:model}
    We briefly describe the modelling of the SMBH growth
    \citep[][for more details]{Shirakata19}.
    The seed BH mass is $10^3 M_\odot$ for all seed BHs
		which are placed when a galaxy newly forms.
    We assume that an SMBH grows with its host bulge via starbursts induced by galaxy mergers and/or disc instabilities.
    The accreted gas mass onto the SMBH, $\Delta M_\mathrm{acc}$,
    and stellar mass formed by a starburst, $\Delta M_\mathrm{star,burst}$, have the following relation:
    \begin{equation}
      \Delta M_\mathrm{acc} = f_\mathrm{BH} \Delta M_\mathrm{star,burst},
    \end{equation}
		where $f_\mathrm{BH} = 0.02$ is chosen to reproduce the local $M_\mathrm{BH}$ -- $M_\mathrm{bulge}$ relation.

		The gas accretion rate is described as follows:
		\begin{equation}
			\dot{M}_\mathrm{BH} (t) = \frac{\Delta M_\mathrm{acc}}{t_\mathrm{acc}} \exp\left(\frac{t-t_\mathrm{start}}{t_\mathrm{acc}}\right), \label{eq:Mdot}
		\end{equation}
		where $t_\mathrm{start}$ and $t_\mathrm{acc}$ are the starting time
		of the accretion and the accretion timescale per one accretion event, respectively.
		We use the same model of the accretion timescale as \cite{Shirakata19},
		$t_\mathrm{acc} = \alpha_\mathrm{bulge} t_\mathrm{dyn,bulge} + t_\mathrm{loss}$.
		The first term of the right-hand side is proportional to the dynamical time of their host bulges,
		$t_\mathrm{dyn,bulge}$, where the values of free parameter, $\alpha_\mathrm{bulge}$, is $0.58$.
		The second term represents the timescale for the angular momentum loss in the ``gas reservoir'' (e.g. circumnuclear discs)
		and accretion disc. We define $t_\mathrm{loss}$ as
		$t_\mathrm{loss,0} (M_\mathrm{BH}/M_\odot)^{\gamma_\mathrm{BH}} (\Delta M_\mathrm{acc}/M_\odot)^{\gamma_\mathrm{gas}}$,
		where the free parameters, $t_\mathrm{loss,0}$, $\gamma_\mathrm{BH}$, and $\gamma_\mathrm{gas}$, are
		1.0 Gyr, 3.5, and -4.0, respectively \citep[see][]{Shirakata19}.

\begin{figure}
      \begin{center}
        \includegraphics[width=\hsize]{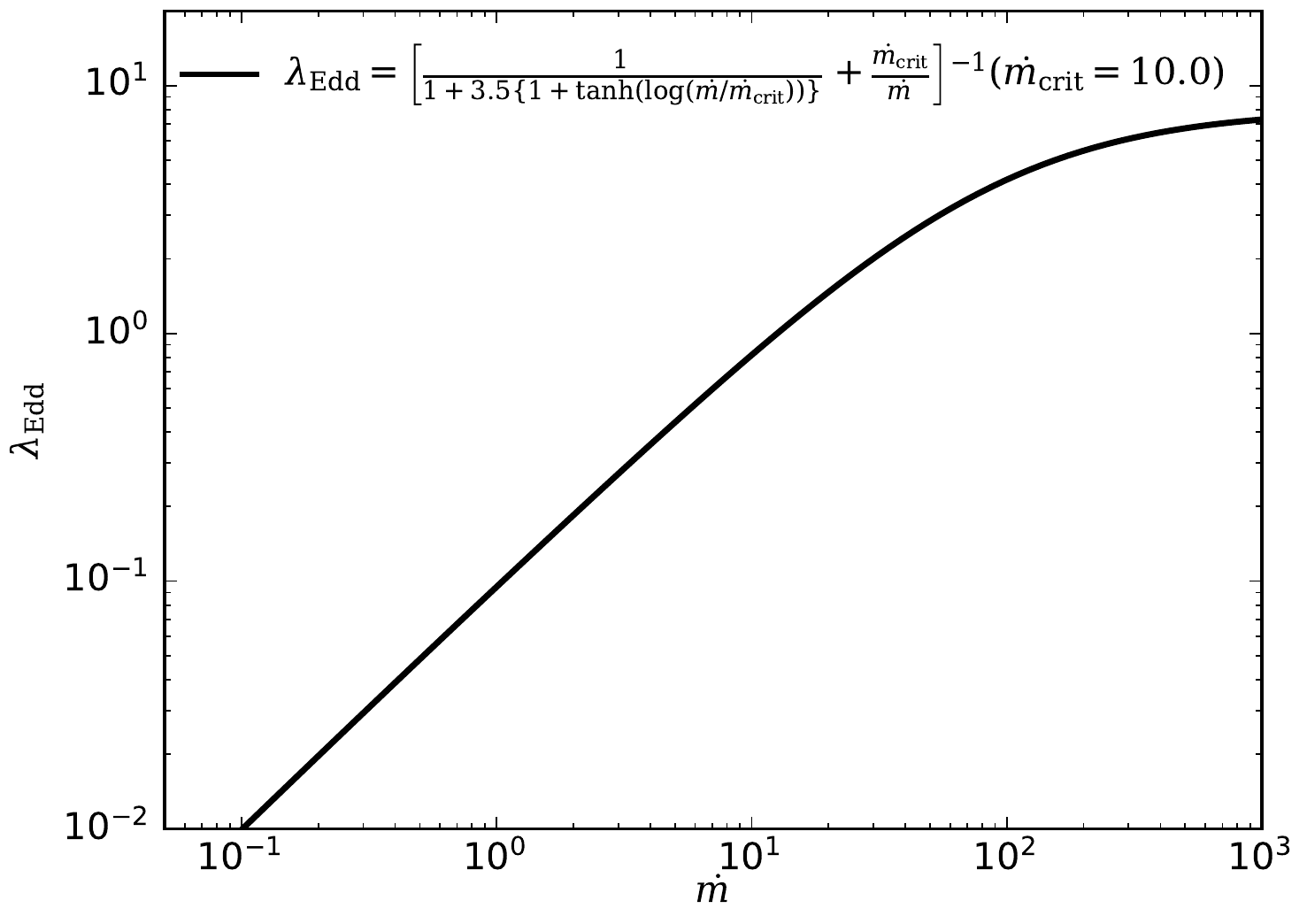}
      \end{center}
      \caption{The conversion relation between $\dot{m}$ and $\lambda_\mathrm{Edd}$
      (Eq. \ref{eq:Lbol}). }
      \label{fig:Mdot2Lbol}
    \end{figure}

		The AGN bolometric luminosity is described with $\dot{m} \equiv \dot{M}_\mathrm{BH}/\dot{M}_\mathrm{Edd}$
    ($\dot{M}_\mathrm{Edd} = L_\mathrm{Edd}/c^2$) as
		\begin{equation}
			\lambda_\mathrm{Edd} = \frac{L_\mathrm{bol}}{L_\mathrm{Edd}} = \left[\frac{1}{1+3.5\{1+\tanh(\log(\dot{m}/\dot{m}_\mathrm{crit}))\}} + \frac{\dot{m}_\mathrm{crit}}{\dot{m}}\right]^{-1}.
			\label{eq:Lbol}
		\end{equation}
		The luminosity of a super-Eddington disc is expected
		to scale in a logarithmic way \citep[e.g.][]{MK00}, in principle.
		Here we employ the formula, Eq. \ref{eq:Lbol}, based on \cite{Kawaguchi03},
		which takes into account various corrections
		(e.g. gravitational redshift, transverse Doppler effect, etc.).
		In this paper, $\dot{m}_\mathrm{crit} = 10$ is assumed.
		The behaviour of the formula is shown in Fig. \ref{fig:Mdot2Lbol}.
		The radiation efficiency, $\epsilon$ is defined as $\lambda_\mathrm{Edd} / \dot{m}$.
		In the super-Eddington regime (i.e. $\dot{m} > \dot{m}_\mathrm{crit}$),
		$\epsilon$ gradually decreases from $0.1$ at $\dot{m} = \dot{m}_\mathrm{crit}$.
		The value of $\epsilon$ is $\sim 0.1$ until $\dot{m} \sim 30$,
		and $\sim 0.01$ at $\dot{m} \sim 600$.
		For deriving the hard $X$-ray (2-10 keV) luminosity of AGNs, $L_X$, we employ
		a bolometric correction obtained by the \cite{Marconi04}.
    For broad-band luminosities in the optical bands, the bolometric correction of AGNs,
    $f \equiv L_\mathrm{bol} / (\nu L_\nu)$,
    is calculated with a template spectral energy distribution of AGNs
    presented in \cite{KSM01}, considering the Doppler shift of the wavelength.
    For $g$- and $i$- bands (the effective wavelengths are 4700 and 7472 \AA, respectively),
    $f = 5.20$ and $7.12$, respectively, at $z \sim 0$.

  \section{The Eddington Ratio Distribution Functions}
    \label{sec:results}
		\subsection{Theoretical Predictions}
\begin{figure*}
        \begin{center}
          \includegraphics[width=\hsize]{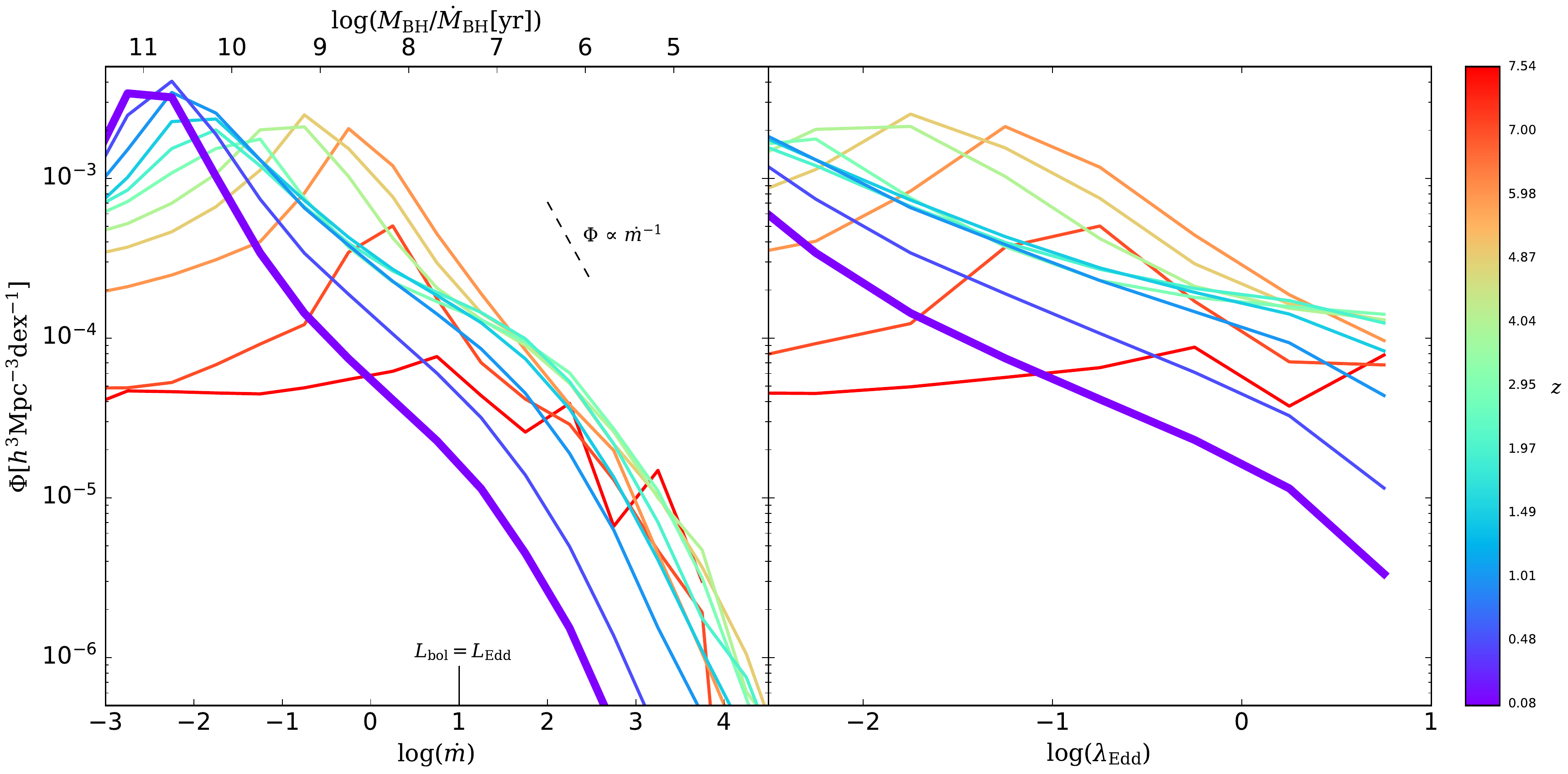}
        \end{center}
        \caption{ERDFs of all AGNs obtained with the fiducial model.
								Left and right panels show the distribution of $\dot{m}$ and $\lambda_\mathrm{Edd}$,
								respectively. Colors describe redshifts ($0.0 < z < 8.0$).
								The black dashed line in the left panel shows the critical slope: $\Phi \propto \dot{m}^{-1}$.
								The thick solid lines describe the ERDF at $z \sim 0.08$.
                The growth timescale, $M_\mathrm{BH}/\dot{M}_\mathrm{BH}$, also appears in the left panel.
								The values of $\Phi$ monotonically increases from $z \sim 0.08$ to $5.98$ with a similar shape
								and decreases at higher redshift with a flatter slope
								at $\log(\dot{m}) = 1$ and $\log(\lambda_\mathrm{Edd}) = 0$.}
        \label{fig:ERDF}
      \end{figure*}

    In Fig.~\ref{fig:ERDF}, we show the ERDFs at $0 < z < 8$
    with the fiducial model.
		In this paper, the ERDF is defined as $\Phi = dn (z, \log{\dot{m}}) / d\log(\dot{m})$,
		or $\Phi = dn (z, \log{\lambda_\mathrm{Edd}}) / d\log(\lambda_\mathrm{Edd})$,
		which is the number density with a fixed $d \log (\dot{m})$ or $d \log (\lambda_\mathrm{Edd})$.
    The left and right panels show the distribution of $\dot{m}$ and $\lambda_\mathrm{Edd}$,
    respectively, for all AGNs (i.e. $M_\mathrm{BH} \geq M_\mathrm{seed}$ and $L_\mathrm{bol} > 0$).
		The bolometric luminosity corresponds to the Eddington luminosity when $\dot{m} \sim \dot{m}_\mathrm{crit}$.
    The dashed line in the left panel indicates a critical slope ($\Phi \propto \dot{m}^{-1}$)
		at which the gas accretion with different $\log (\dot{m})$ contributes to SMBH growth equally
    (i.e. $\Phi~\dot{m}~\Delta \log(\dot{m}) = const.$).
		The thick solid lines describe the ERDFs at $z \sim 0.08$.
		The values of $\Phi$ monotonically increases from $z \sim 0.08$ to $5.98$ with a similar shape
		and decreases at higher redshift with a flatter slope
		at $\log(\dot{m}) = 1$ and $\log(\lambda_\mathrm{Edd}) = 0$.
		The slope of the ERDFs becomes flatter at higher redshift,
		meaning that the relative number of AGNs with higher Eddington ratio increases at higher redshift.
    Upper label in Fig. \ref{fig:ERDF} shows SMBH growth timescale, $M_\mathrm{BH}/\dot{M}_\mathrm{BH}$.
    If the growth timescale is shorter than the age of the Universe at a given redshift,
    then the SMBH can acquire their mass by gas accretion
    until that redshift without requiring heavy (i.e. $\gtrsim 10^5 M_\odot$) seed BH.
    Thanks to the super-Eddington accretion, SMBHs can grow much faster than
    $M_\mathrm{BH}/(10 \dot{M}_\mathrm{Edd}) \sim 45$ Myr (i.e. Salpeter timescale with $\epsilon \sim 0.1$),
		which is much less than the age of the Universe at which the SMBHs exist.

\begin{figure}
        \begin{center}
          \includegraphics[width=\hsize]{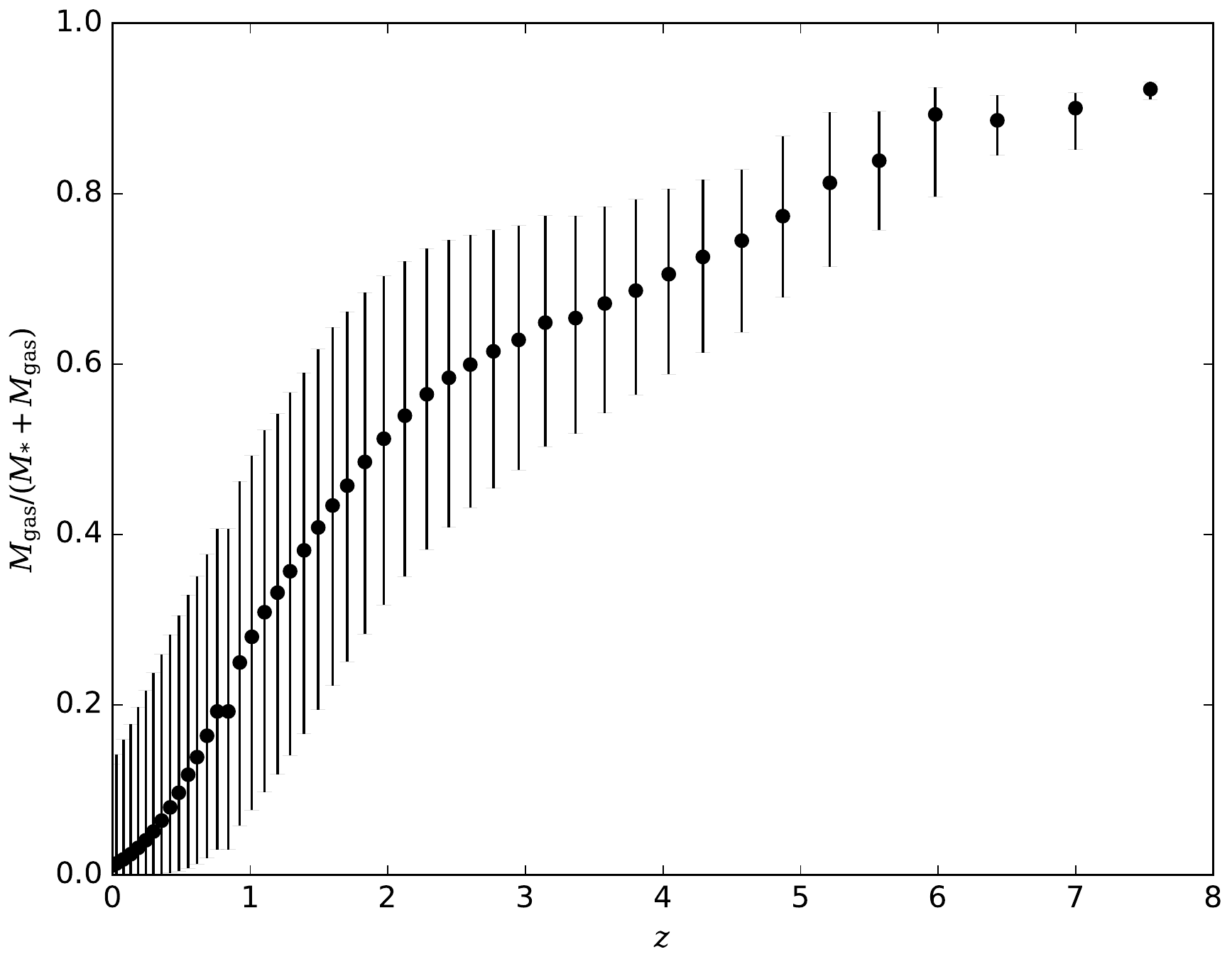}
        \end{center}
        \caption{The redshift evolution of the gas fraction of AGN host galaxies obtained by the model.
        Black points show the median value at a redshift and error bars mean the 25 and 75 percentiles of the scatter.}
        \label{fig:GasFrac}
      \end{figure}

      The reason why SMBH growths at higher redshift have higher Eddington ratio is shown in Fig. \ref{fig:GasFrac}.
      The figure shows the evolution of the gas fraction of AGN host galaxies, defined as $M_\mathrm{gas} / (M_\mathrm{gas} + M_*)$,
      where $M_\mathrm{gas}$ and $M_*$ are the cold gas and stellar masses of a galaxy (disc $+$ bulge), respectively.
      The gas fraction is higher at higher redshift, e.g. $\sim 0.8$ at $z \sim 5$.
      A higher gas fraction leads to a high $\dot{M}_\mathrm{BH}$, in general.
      Therefore, the Eddington ratio tends to be higher at higher redshift even if the SMBH mass is the same.

\begin{figure*}
				\begin{center}
					\includegraphics[width=\hsize]{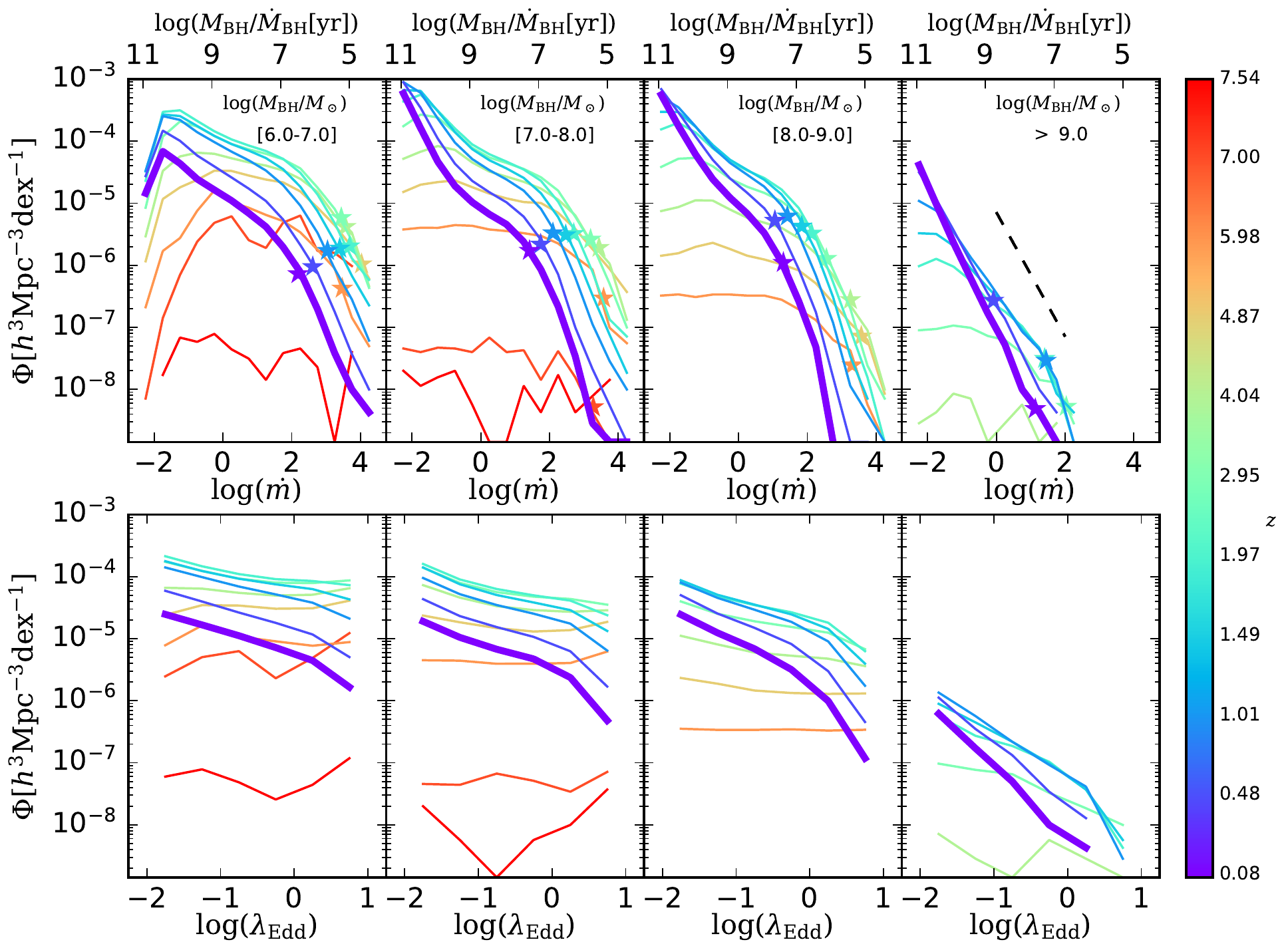}
				\end{center}
				\caption{ERDFs of AGNs with $L_X > 10^{41}$ erg/s obtained with the fiducial model.
								Top and bottom panels show the distribution of $\dot{m}$ and $\lambda_\mathrm{Edd}$,
								respectively. Colors describe redshifts ($0.0 < z < 8.0$).
								We create four subsamples by $M_\mathrm{BH}$; $\log(M_\mathrm{BH}/M_\odot)$ is [6,7], [7,8], [8,9], and $> 9$,
								from left to right panels.
                The black dashed line in the right panel shows the slope $\Phi \propto (\dot{m})^{-1}$
                and stars show the point at which the slope becomes $-1$ and $\dot{m} \Phi$ takes the largest value.
                The growth timescale, $M_\mathrm{BH}/\dot{M}_\mathrm{BH}$, also appears in the top axis.
								The thick solid lines describe the ERDF at $z \sim 0.08$.
								The values of $\Phi$ for subsample with $\log(M_\mathrm{BH}) = [6-7], [7-8], [8.9]$, and $> 9$
								monotonically increases from $z \sim 0.08$ to $2.95$, $2.95$, $1.97$, and $1.49$, respectively,
								and decreases at higher redshifts
								at $\log(\dot{m}) = 1$ and $\log(\lambda_\mathrm{Edd}) = 0$.}
        \label{fig:ERDF_bin}
    \end{figure*}

		To present the mass dependence on the ERDFs, we
		show the ERDFs of AGNs with $L_X > 10^{41}$ erg/s for different SMBH mass bins,
    $\log(M_\mathrm{BH}/M_\odot) = [6,7]$, $[7,8]$, $[8,9]$, and $>9$, respectively, in Fig. \ref{fig:ERDF_bin}.
    The colour of each line indicates the redshift, $0 < z < 8$, as shown in the colour bar.
		The thick solid lines describe the ERDFs at $z \sim 0.08$.
		The values of $\Phi$ for subsample with $\log(M_\mathrm{BH}) = [6,7], [7,8], [8,9]$, and $> 9$
		monotonically increases with a similar shape
		from $z \sim 0.08$ to $2.95$, $2.95$, $1.97$, and $1.49$, respectively,
		and decreases at higher redshifts with a flatter slope
		at $\log(\dot{m}) = 1$ and $\log(\lambda_\mathrm{Edd}) = 0$.
    Our calculation shows that ERDFs, not only at higher redshift
		but also with less massive BHs tend to have flatter slopes,
    indicating that present day SMBHs are formed via super-Eddington accretion at higher redshifts.

		\begin{figure}
				\begin{center}
					\includegraphics[width=\hsize]{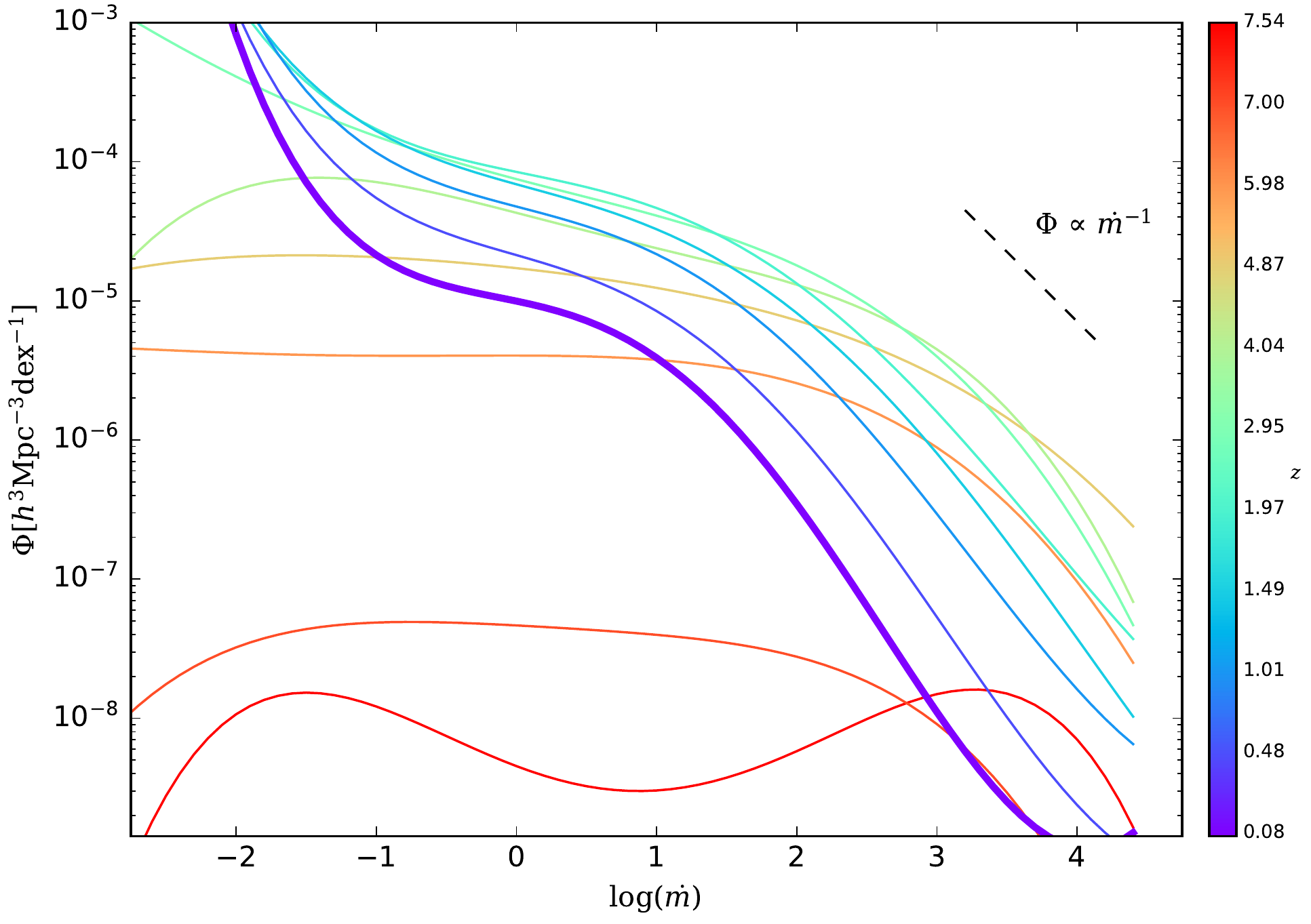}
				\end{center}
        \caption{The ERDF fitting result in $\log(M_\mathrm{BH} [M_\odot]) = $ [7,8].}
					\label{fig:ERDF_fit}
		\end{figure}

		We then define $\dot{m}_\mathrm{knee}$ as $\dot{m}$
		at which the ERDF has the critical slope ($-1$) and $\dot{m} \Phi$ has the largest value.
		Accretion with $\dot{m} = \dot{m}_\mathrm{knee}$ contributes to the SMBH growth
		most at that redshift and at that $M_\mathrm{BH}$ range.
		We obtain the points of contact by fitting ERDFs of $\dot{m}$ with fourth-order functions
		at $-2 < \log(\dot{m}) < 4$.
		An example of the fitting results is shown in Fig. \ref{fig:ERDF_fit} for $M_\mathrm{BH} = 10^{7-8} M_\odot$.
		Because of the difficulties for obtaining smooth curves of ERDFs,
		the fitting result has a larger uncertainty at higher redshift.

\begin{figure*}
        \begin{center}
          \includegraphics[width=\hsize]{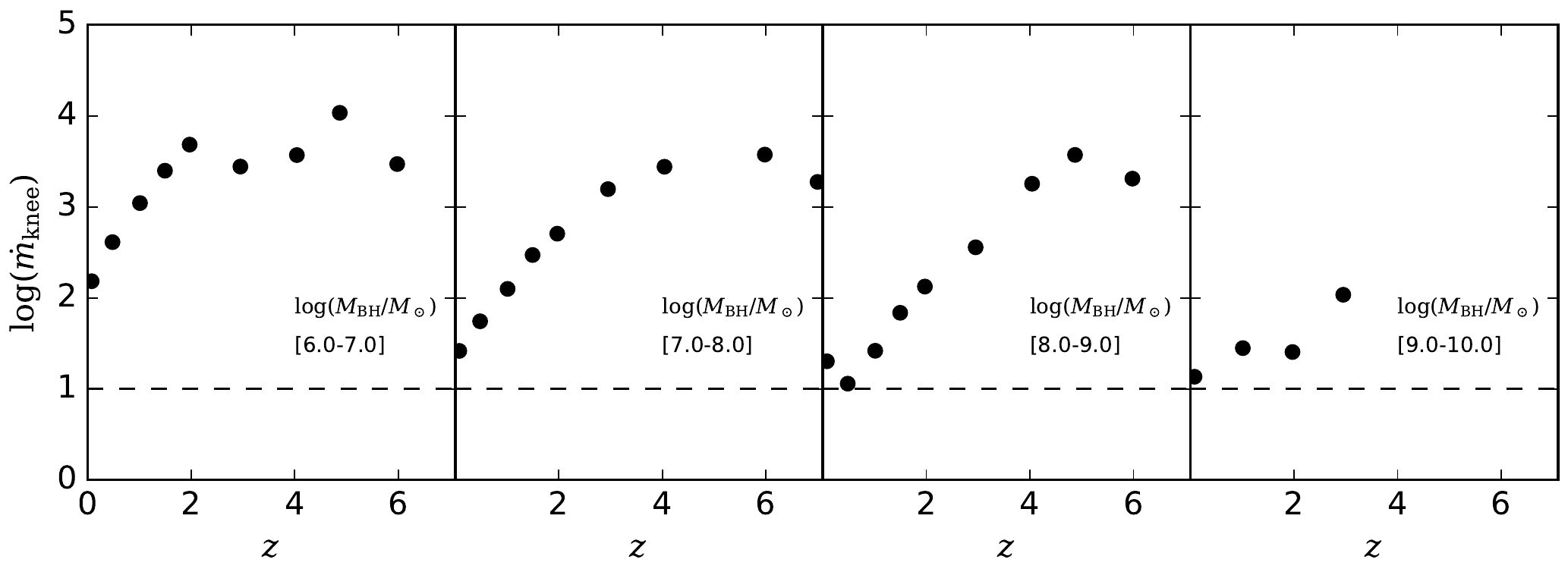}
        \end{center}
        \caption{The evolution of the $\dot{m}$ at which the slope of ERDF becomes $\sim -1$.
					Four panels show the results with the different $M_\mathrm{BH}$ bins,
          and are shown as stars in Fig. \protect\ref{fig:ERDF_bin}.
          Dashed lines indicate $\dot{m}_\mathrm{knee} = \dot{m}_\mathrm{crit}$.
				  }
          \label{fig:ERDF-knee}
    \end{figure*}

		Fig. \ref{fig:ERDF-knee} shows the relation between $\dot{m}_\mathrm{knee}$ and redshift
		of the same subsample as Fig. \ref{fig:ERDF_bin}.
		In all SMBH mass ranges, $\dot{m}_\mathrm{knee}$ increases towards higher redshift.
		Besides, $\dot{m}_\mathrm{knee}$ is larger for less massive SMBHs when we compare AGNs at a fixed redshift.
		These $\dot{m}_\mathrm{knee}$ evolutions in redshift and mass indicate that
		super-Eddington accretion is more common for less massive SMBHs and AGNs at higher redshift.
    These results of the Eddington ratio evolution are consistent with another approach to constraining
    the SMBH growth by integrating the AGN LFs and
    the SMBH MFs \citep{Soltan82,CT92},
		showing that the sub-Eddington accretion is responsible for SMBH growth at $M_\mathrm{BH} \gtrsim 10^9 M_\odot$
    \citep[e.g.,][]{YT02,FI99}.
    For lower SMBH masses and higher redshifts, in contrast, our model shows that the super-Eddington accretion
    is the primary growth mode of SMBHs.

\begin{figure}
      \begin{center}
        \includegraphics[width=\hsize]{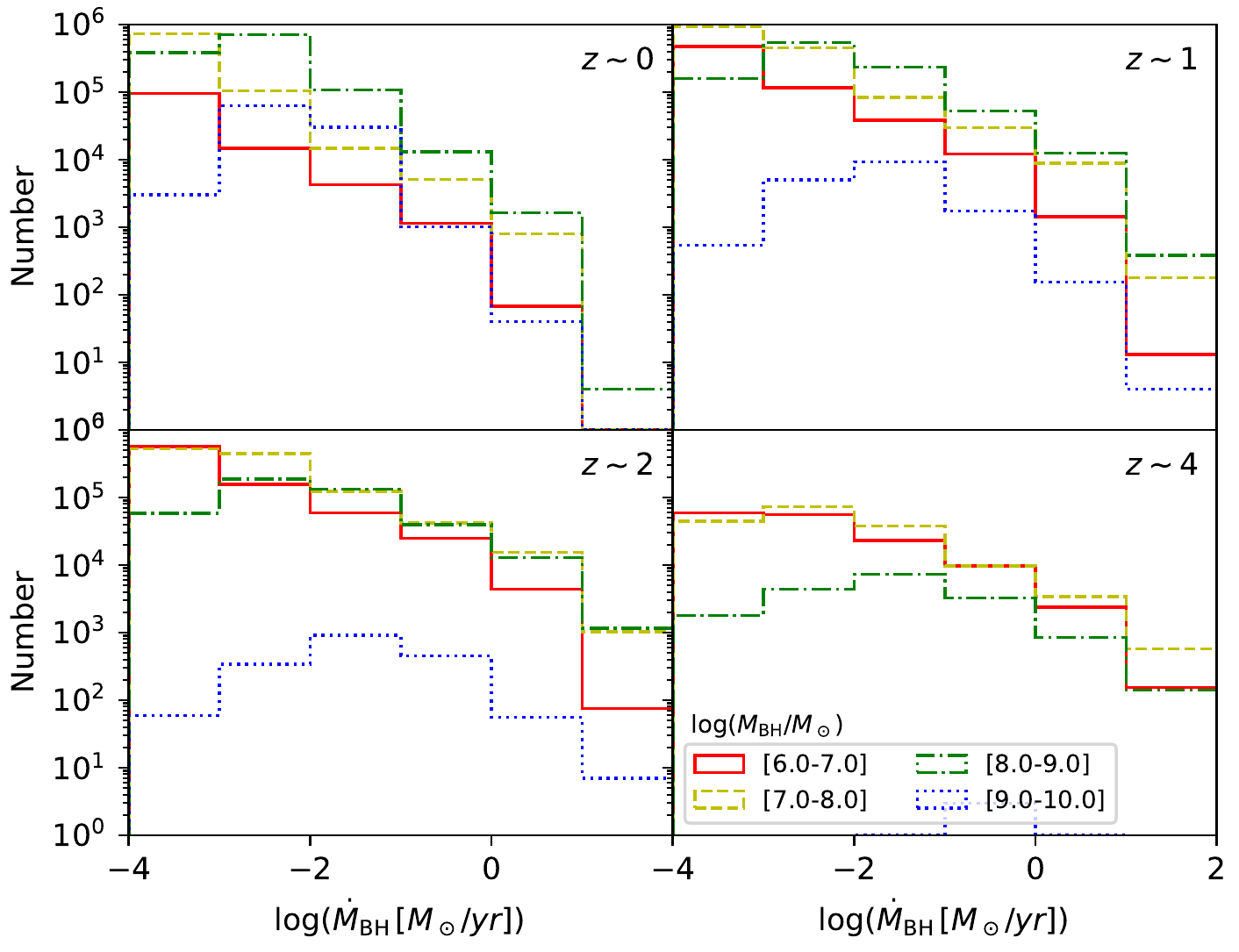}
      \end{center}
			\caption{The number distributions of $\dot{M}_\mathrm{BH}$ with different SMBH mass ranges
			at $z = 0, 1,2,$ and $4$.}
      \label{fig:Mdot_hist}
    \end{figure}

		The dependence of ERDFs on the SMBH mass comes from the denominator of $\dot{m}$, i.e. $\dot{M}_\mathrm{Edd}$ itself,
		which is proportional to the SMBH mass.
    Fig. \ref{fig:Mdot_hist} shows the distribution of $\dot{M}_\mathrm{BH}$ with four
    SMBH mass bins at $z \sim 0, 1, 2,$ and $4$.
    The accretion rate, $\dot{M}_\mathrm{BH}$, which is not normalised by the Eddington accretion rate,
    has no clear trends with the SMBH mass.
    The distribution of $\dot{M}_\mathrm{BH}$ and $\dot{m}$ means that
    the accretion rate is not regulated by the SMBH mass nor the Eddington limit.
    We note that the accretion rate irrespective of the SMBH mass nor the Eddington limit
    is also supported observationally \citep{CK04}, where the maximum accretion rate
    of AGNs with different SMBH masses are similar at $z \sim 0$.

\begin{figure}
      \begin{center}
        \includegraphics[width=\hsize]{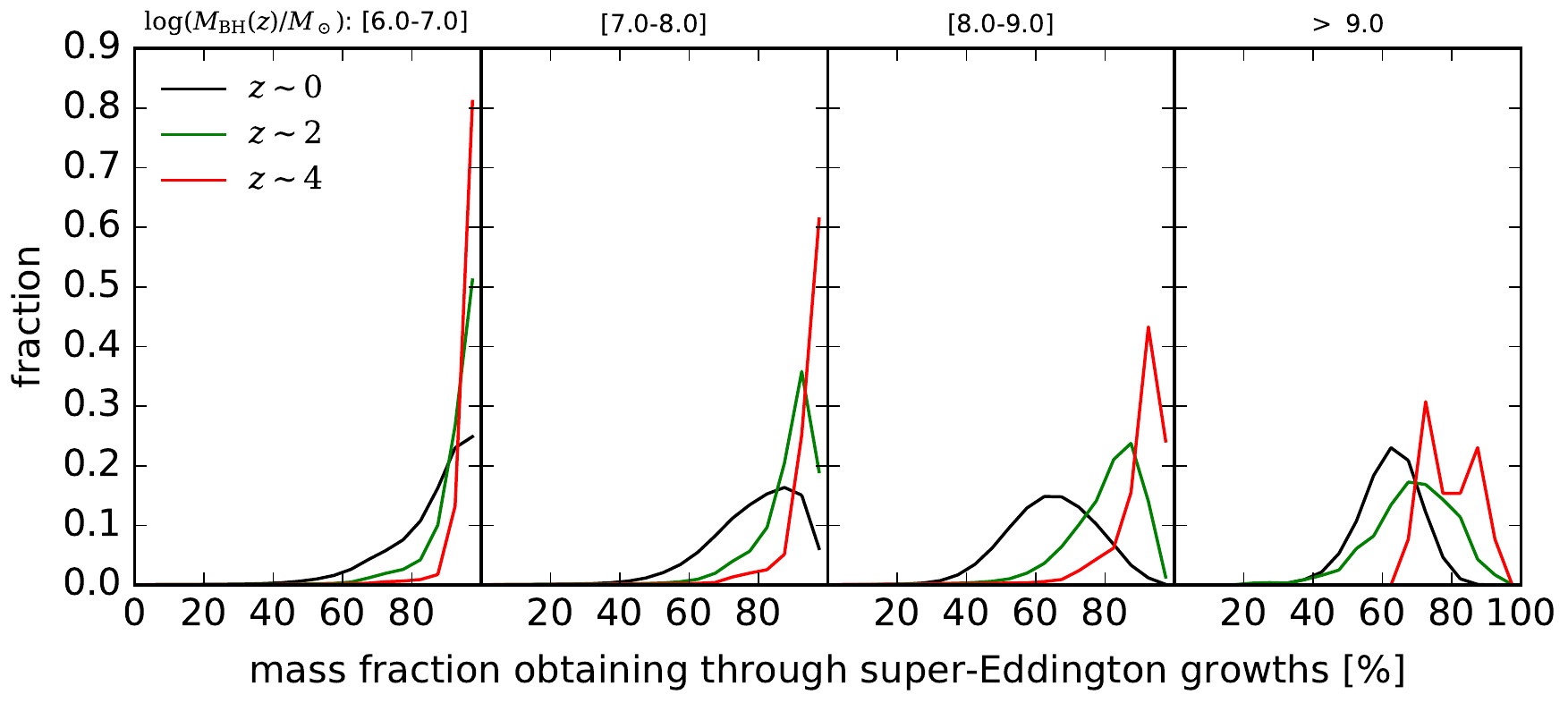}
      \end{center}
			\caption{Distribution of the fraction of the mass
			obtained by super-Eddington accretion out of the total SMBH mass,
			for different mass ranges and different redshift.
			For a given SMBH mass range, super-Eddington growth becomes more dominant
			at higher redshift. At each redshift, SMBHs with smaller masses
			tend to have larger fractions of the mass accumulated via super-Eddington growth.}
      \label{fig:SE-hist}
    \end{figure}

		In this section, we have shown that the number fraction of
		AGNs with super-Eddington growth is higher at higher redshift
		or for less massive SMBHs, as suggested from the slope of ERDFs.
		Here, we show this trend in a different view.
		Each SMBH has a certain relative fraction of
		the mass acquired via super-Eddington growth
		out of the total SMBH mass.
		Fig \ref{fig:SE-hist} shows the distribution of such mass fraction,
		for SMBHs at different mass ranges and different redshift.
		As seen in Figs. \ref{fig:ERDF}, \ref{fig:ERDF_bin}, and \ref{fig:ERDF-knee}, 
		super-Eddington growth becomes more dominant at higher redshift
		for a given SMBH mass range.
		Also, for a given redshift,
		SMBHs with smaller masses have larger fractions of the mass
		obtained by super-Eddington accretion.
		Further analysis, in the context of the \Soltan argument, will be presented in a subsequent paper.

		\subsection{Comparisons with Observations}
		\label{sec:obscomp}
		Here we compare our results with observations.
    We note that we do not assume the obscuration of AGNs
		for simplicity: we assume the obscured fraction is zero
    since it could depend on the bolometric luminosity and the Eddington ratio itself
		\citep[e.g.][]{Lusso12} and the dependencies are unclear.
		The equivalent comparisons with observations cannot be made now,
		and we leave them for future studies.

\begin{figure*}
        \begin{center}
          \includegraphics[width=\hsize]{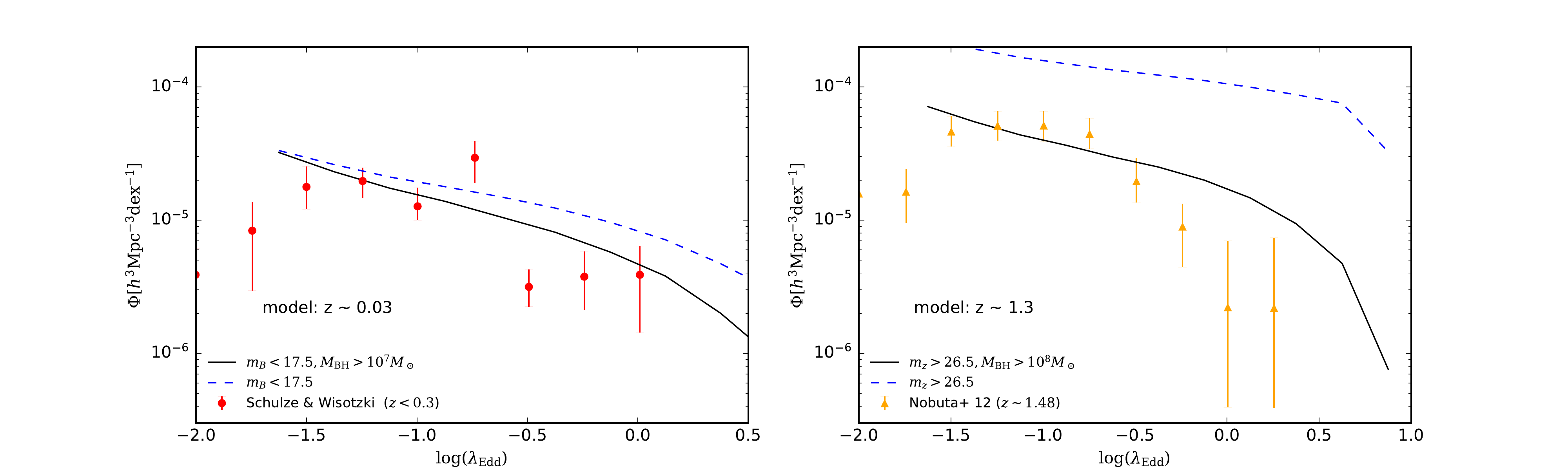}
        \end{center}
        \caption{Comparisons of ERDFs with observations at $z < 0.3$ \protect\citep[left panel, red circles]{SW10} and
        at $z \sim 1.4$ \protect\citep[][right panel, orange triangles]{Nobuta12}.
        We select AGNs with the similar sample selection to each observation (black solid lines), where $m_B$ and $m_z$ are
        the apparent magnitude of $B$- and $z$- bands.
        We also show the results without imposing any limits for SMBH mass (blue dashed lines).}
          \label{fig:comp_obs}
    \end{figure*}

    Fig. \ref{fig:comp_obs} shows the comparison of the fiducial model
    with observations of the ERDF \citep{SW10, Nobuta12}.
    We select AGNs with a similar sample selection to each observational study (black lines).
		In this case, the model ERDF is roughly consistent with observations.
    Without the SMBH mass limit, however, the slope of the model ERDFs become flatter
    than observed since the accretion rate of less massive SMBHs in the model can easily reach and exceed the Eddington limit.
    The difference of the slope becomes larger at higher redshift, which suggests the importance of deeper observations
    (resulting in a weaker mass limit) to obtain ERDFs more accurately.

\begin{figure*}
        \begin{center}
          \includegraphics[width=\hsize]{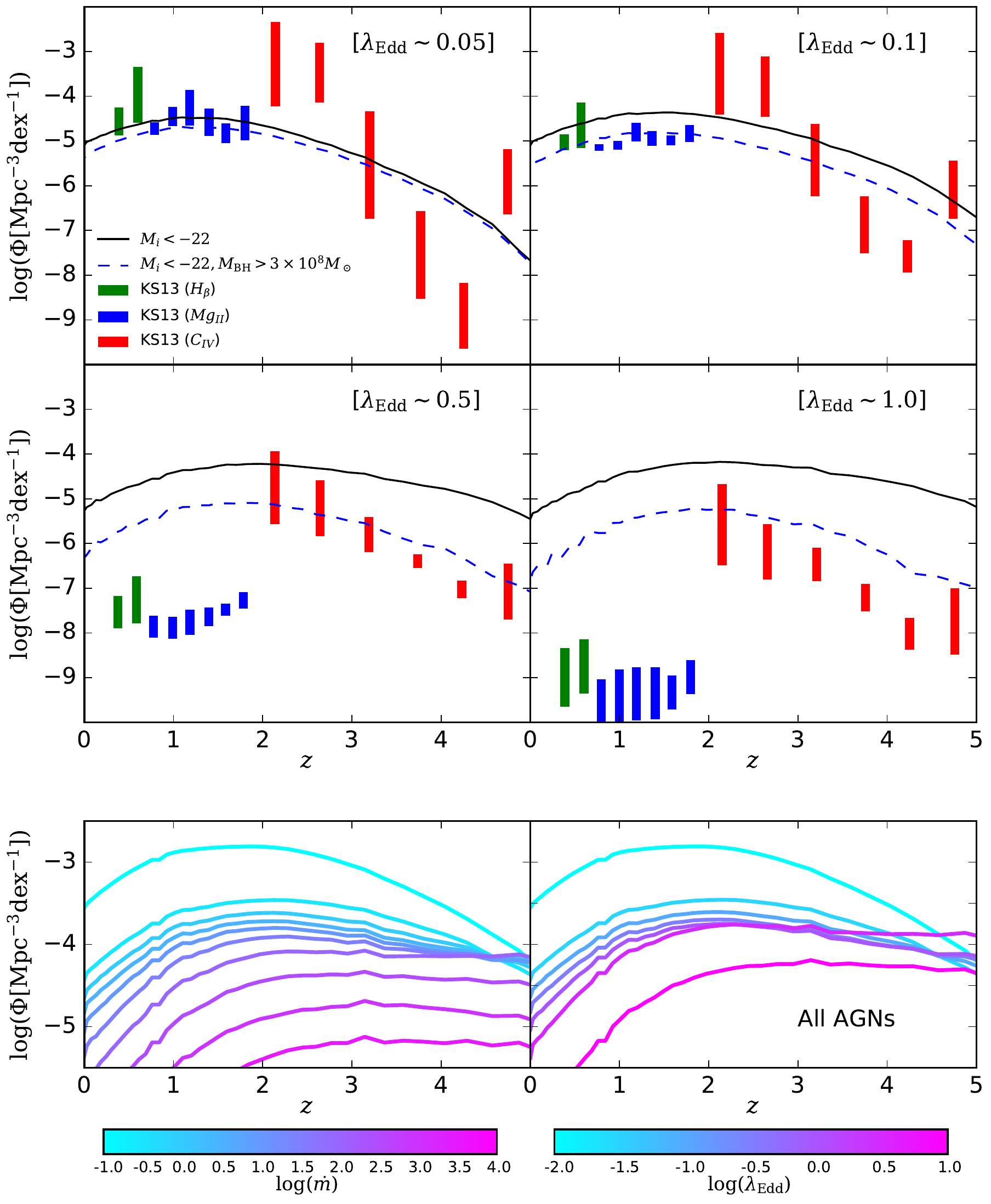}
        \end{center}
        \caption{The redshift evolution of the number density of AGNs for different $\lambda_\mathrm{Edd}$ ranges at $0 < z < 5$.
					\protect\textit{Top four panels}: Comparisons with the model results with the observation (\protect\citetalias{KS13}).
          The black solid and blue dashed lines show the model results for AGNs with $M_i < -22$ and $M_i < -22$
          and $M_\mathrm{BH} > 10^{8.5} M_\odot$, respectively.
          The latter ones are similar sample selection as \protect\citetalias{KS13}.
          Green, blue, and red bars indicate results with \protect\citetalias{KS13}, whose colour difference shows the
          results with different emission lines for estimating $M_\mathrm{BH}$ ($H_\beta, Mg_{II}$, and $C_{IV}$, respectively).
          \protect\textit{Bottom panels}: The redshift evolution of the number density with the fiducial model.
          We select all model AGNs with $L_X > 10^{41}$ erg/s. The colour indicates
					$\log(\dot{m})$ from $-1$ to $4$ in the left panel and
					$\log(\lambda_\mathrm{Edd})$ from $-2$ to $1$ in the right panel, as shown in colour bars.}
          \label{fig:ERdens}
    \end{figure*}

		Fig. \ref{fig:ERdens} shows the evolution of the number density of AGNs for different $\lambda_\mathrm{Edd}$ ranges.
		The model results for AGNs with $M_i < -22$ and those with $M_i < -22$ and $M_\mathrm{BH} > 10^{8.5} M_\odot$ are shown
		in the upper four panels, where $M_i$ is the absolute magnitude in $i$-band.
		The latter selection is similar to that of \citetalias{KS13}, based on SDSS data.
		Our results, especially with the latter sample selection, are consistent with
		the estimates of \citetalias{KS13}.
		However, the results of \citetalias{KS13} have a discontinuous shape,
		which is likely to be originated from
		the difference of the emission lines used for the $M_\mathrm{BH}$ estimation.
		Since their typical equivalent-widths are different from line to line,
		the effective depth of spectroscopic surveys changes
		when one changes the emission line to use.
		In other words, observational determination of ERDFs at high redshift is very uncertain and
		is not yet ready for comparison with models.
    Deeper surveys such as those with the Subaru Prime Focus Spectrograph \citep{Takada14} are needed.
		We find that there is a trend that higher $\lambda_\mathrm{Edd}$ ranges present their peak number density
		at higher redshift, which is qualitatively consistent with \citetalias{KS13}.
		Namely, the more rapid growth occurres at higher redshift,
		indicating the slowing down, decelerating or slacking of cosmic SMBH growth.
		This trend can be seen in the bottom panels of Fig. \ref{fig:ERdens},
		in which we present the redshift evolution of the number density of AGNs
		for each $\dot{m}$ and $\lambda_\mathrm{Edd}$ (left and right panels, respectively)
		indicated by the colour of lines.
		In these panels, we select all AGNs with $L_X > 10^{41}$ erg/s.
		While the number density of the higher Eddington ratio remains constant at $z \gtrsim 3$
		and decreases at lower redshift,
		that of the lower Eddington ratio gradually increases towards lower redshift,
		clearly indicating the slowing down of cosmic SMBH growth.

		As shown in Fig. \ref{fig:ERDF_bin}, our model predicts that
		super-Eddington accretion becomes more important at higher redshift and for relatively lesser massive SMBHs.
		Observational determinations of ERDFs tend to underestimate
		the number density of super-Eddington accreting AGNs,
		whose optical flux is under the observational limit and/or whose SMBH mass has not been evaluated,
		and undervalue the importance of the super-Eddington growth (see also Fig. \ref{fig:comp_obs}).

\begin{figure}
        \begin{center}
          \includegraphics[width=0.9\hsize]{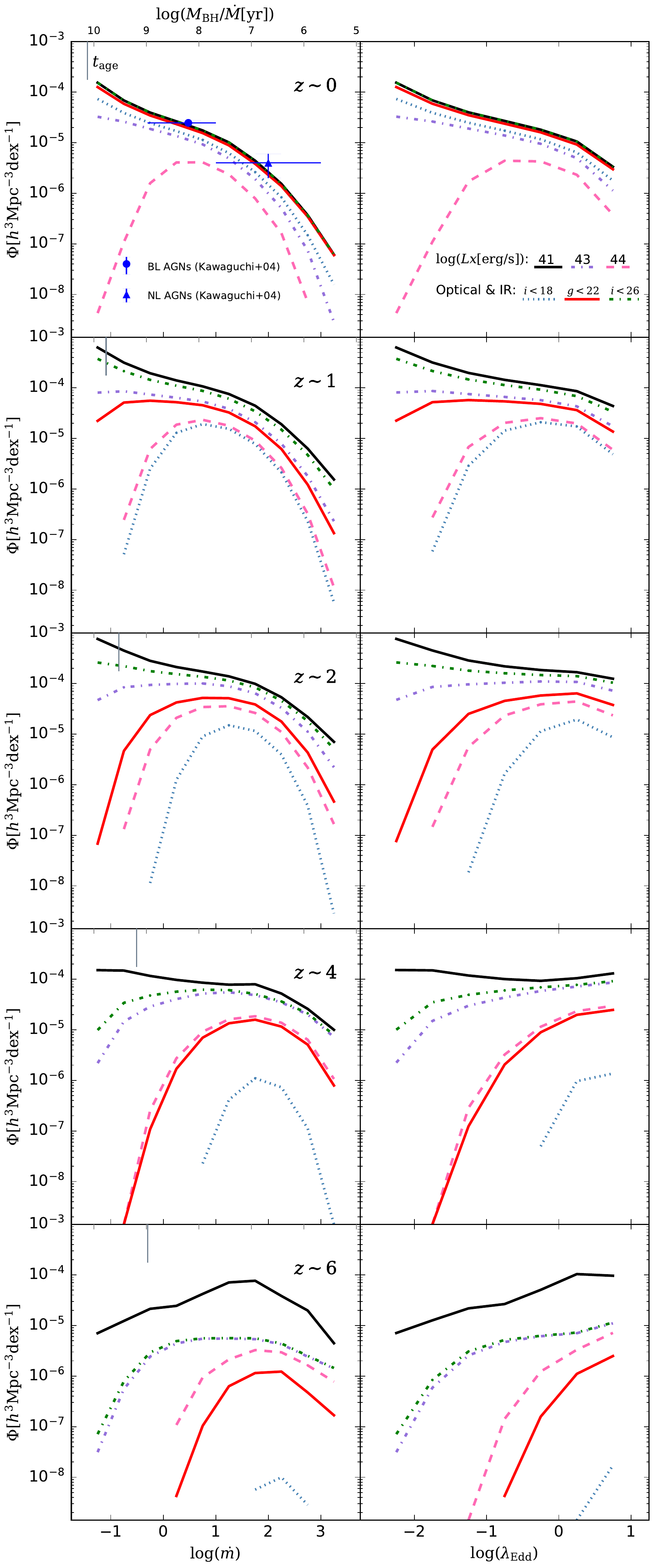}
        \end{center}
        \caption{The ERDF with varying the sample selection at $0 < z < 6$.
        Left and right panels show the distribution functions of $\dot{m}$ and $\lambda_\mathrm{Edd}$, respectively.
        We employ different absolute luminosity cuts with $X$-ray, $L_X > 10^{41}$, $> 10^{43}$, and $> 10^{44}$ erg/s
        (black solid, purple dot-dashed, and pink dashed lines), and optical flux cuts, $i < 18$, $g < 22$, and $i < 25.9$
        (blue dotted, red solid, and green dot-dashed lines). The optical flux cuts correspond to those of SDSS-DR7
        \protect\citep{Schneider10}, SDSS-DR12 \protect\citep{Paris17}, and forthcoming Subaru HSC-SSP survey catalogue with
        wide layers \protect\citep{Aihara17}.
        We overplot the result obtained by \protect\cite{Kawaguchi04June} for the relative fraction of broad and narrow line AGNs at $z < 0.5$.
				We also indicate the age of the Universe at each redshift, $t_\mathrm{age}$.
    SMBHs, whose growth timescale ($M_\mathrm{BH}/\dot{M}_\mathrm{BH}$) is shorter than $t_\mathrm{age}$,
		can acquire their mass by the current $\dot{M}_\mathrm{BH}$
		and do not require heavy seed BHs such as $\gtrsim 10^5 M_\odot$.}
      \label{fig:ERDFsample}
    \end{figure}

    Fig. \ref{fig:ERDFsample} shows the ERDF at $0 < z < 6$
    varying the sample selection.
    We show the results with different absolute luminosity cuts, $L_X > 10^{41}$, $> 10^{43}$, and $> 10^{44}$ erg/s,
    and flux cuts, $i < 18$ (SDSS-DR7 catalogue; \citealt{Schneider10}), $g < 22$ (SDSS-DR10 catalogue; \citealt{Paris17}),
		and $i < 25.9$ (Subaru HSC-SSP wide layer; \citealt{Aihara17}).
    We include all AGNs which satisfy the above cuts
    without considering the type-2 fraction or obscured fraction.
    We note that the model predicts the consistent ratio of local sub- and super-Eddington AGNs
    suggested by various low-$z$ observations \citep{Kawaguchi04June}.
    In \cite{Kawaguchi04June},
		ranges for $\dot{m}$ are estimated from the disc model by Fig. 11 of \cite{Kawaguchi03},
		resulting in $80^{+10}_{-10}$ \% objects in $\dot{m} = 3^{+7}_{-2.5}$,
		and $20^{+10}_{-10}$ \% in $\dot{m} = 100^{+900}_{-90}$.
		When the shallower sample selections are applied, the number density of the AGNs with smaller Eddington ratios
		decreases. Also, the slope and the normalisation at a higher Eddington regime are affected by the sample selection
		since our model predicts that super-Eddington growth is more common for less massive SMBHs.
		The effect of the sample selection becomes more critical at higher redshift.
		The results imply that spectroscopic follow up observations and SMBH mass measurements
		including less luminous AGNs play a crucial role in obtaining nearly ``complete'' ERDFs
    and understanding the cosmic growth of SMBHs.

		The shape of ERDFs with a luminosity/flux cut would change by the bolometric correction.
		We employ the bolometric correction independent from the Eddington ratio and SMBH mass \citep{Marconi04},
		although \cite{JWD12} suggest that it depends on the Eddington ratio and SMBH mass.
		With the type-1 AGN sample at $z < 0.4$, \cite{JWD12} also suggest that
		more luminous AGNs tend to radiate a smaller amount of optical and $X$-ray light,
		which is the opposit trend to that presented in \cite{Marconi04}.
		When we employ the $\lambda_\mathrm{Edd}$-dependency of
		the bolometric correction of \cite{JWD12}, model ERDFs with a luminosity/flux cut
		will become steeper at high $\lambda_\mathrm{Edd}$ regime and become more consistent with current observations.
		Since the bolometric correction has large uncirtainty, we simply employ \cite{Marconi04} in this paper.

  \section{Discussion}
  \label{sec:discussion}
	In Sec. \ref{sec:obscomp}, the shape of the ERDFs with a luminosity/flux cut
	would change by the bolometric correction.
	The shape could also change by the effect of winds/outflows with super-Eddington accretions.
	\cite{Takeo18} perform two dimensional radiation hydrodynamic simulations of gas accretions
	for BHs with $10^{3-7} M_\odot$ and with $5000 \dot{M}_\mathrm{Edd}$,
	whose initial gas distribution is uniform and the gas is in the neutral state.
	For BHs with $> 5 \times 10^5 M_\odot$, the radiation by the gas accretion heats up
	the surrounding gas to $\sim 8000 K$ due to the recombination by hydrogen.
	\cite{Takeo18} conclude that in this case, the outflow rate becomes $\sim 10$ \% of the
	mass accretion rate.
	\cite{JSD17} employ the three dimensional radiation magneto-hydrodynamical simulations
	and find that the outflow rate decreases $15-50$ \% of the mass accretion rate for a BH with $5 \times 10^8 M_\odot$
	and with $250-1500 \dot{M}_\mathrm{Edd}$.
	Although the shape of ERDFs would change by the winds,
	the effects are expected not to be significant and beyond the scope of this paper.
	We leave the introduction of the wind effect for future studies for diminishing the degree of freedom.

	Previous studies with SA models have also presented the ERDFs.
	\cite{Fanidakis12} give the evolution of ERDFs with $M_\mathrm{BH} > 10^6 M_\odot$.
	Their ERDFs show almost no evolution of the slope for the high Eddington ratio,
	in contrast to our study.
	\cite{Hirschmann12,Hirschmann14}, on the other hand, predict flatter ERDFs at higher redshift,
	which is broadly consistent with our result.
	In \cite{Hirschmann12, Hirschmann14}, they employ constant ($= 0.01$) Eddington ratio for AGNs triggered by disc instabilities
	and an Eddington ratio distribution for AGNs triggered by mergers obtained by merger simulations,
	where an AGN host galaxy with a higher cold gas fraction produces an AGN with a higher Eddington ratio.
	Both \cite{Fanidakis12} and \cite{Hirschmann14} show the bimodal distribution of ERDFs,
	which are caused by the different accretion disc properties (i.e. advection-dominated accretion flow and standard disc)
	and/or different accretion modes (i.e. hot-halo mode and starburst/QSO mode).
	The evolution of the peak in the ERDF in the standard disc regime is broadly consistent with our results;
	the peak move to lower Eddington ratio at lower redshift.
	Since the advection-dominated accretion flow regime is out of the observable regime
	and the bolometric correction in this regime is unclear,
	we leave the detailed analysis in this regime for future studies.
	We note that the model of the gas accretion is different in different SA models.
	To discriminate the models, ERDFs, which will be obtained by future observations, are required.

	In Sec. \ref{sec:results}, we have shown that the shape of ERDFs depends on the redshift and mass of SMBHs.
	The mass dependence of the ERDFs is still unclear in observations.
	Observationally, there have been some studies about the relation between
	the Eddington ratio and properties of AGN host galaxies,
	\footnote{Since the ``true'' Eddington ratio is difficult to estimate,
		the AGN luminosity ($X$-ray luminosity, especially)
		divided by the stellar mass of the host galaxy is sometimes used as a proxy for the Eddington ratio.}
		which are also essential for understanding the cosmic growth and coevolution
		of SMBHs and their host galaxies.
	As an example, \cite{Bongiorno16} obtain $L_X/M_*$ and its distribution instead of the Eddington ratio.
	They argue that the shape of the distribution is independent of the stellar mass
	although the normalisation is different.
		Although \cite{Bongiorno16} find no dependence of the shape of ERDFs on the stellar mass,
	their results are obtained with the assumption of $M_\mathrm{BH} = 0.002 M_*$ at $z < 2.5$.
	This assumption is reasonable and can be compared with our model results for local bulge-dominated galaxies.
	However, since the host galaxies of AGNs have various morphologies \citep[e.g.][]{Bruce15}
	and the fraction of disc-dominated galaxies becomes higher with less massive galaxies \citep[e.g.][]{Lang14},
	the mass of SMBHs could be smaller than $0.002 M_*$ especially for less massive galaxies at higher redshift.
	Therefore, the assumption of the mass ratio between SMBHs and their host galaxies
	can affect the mass dependence on the ERDFs.

\begin{figure}
      \begin{center}
        \includegraphics[width=\hsize]{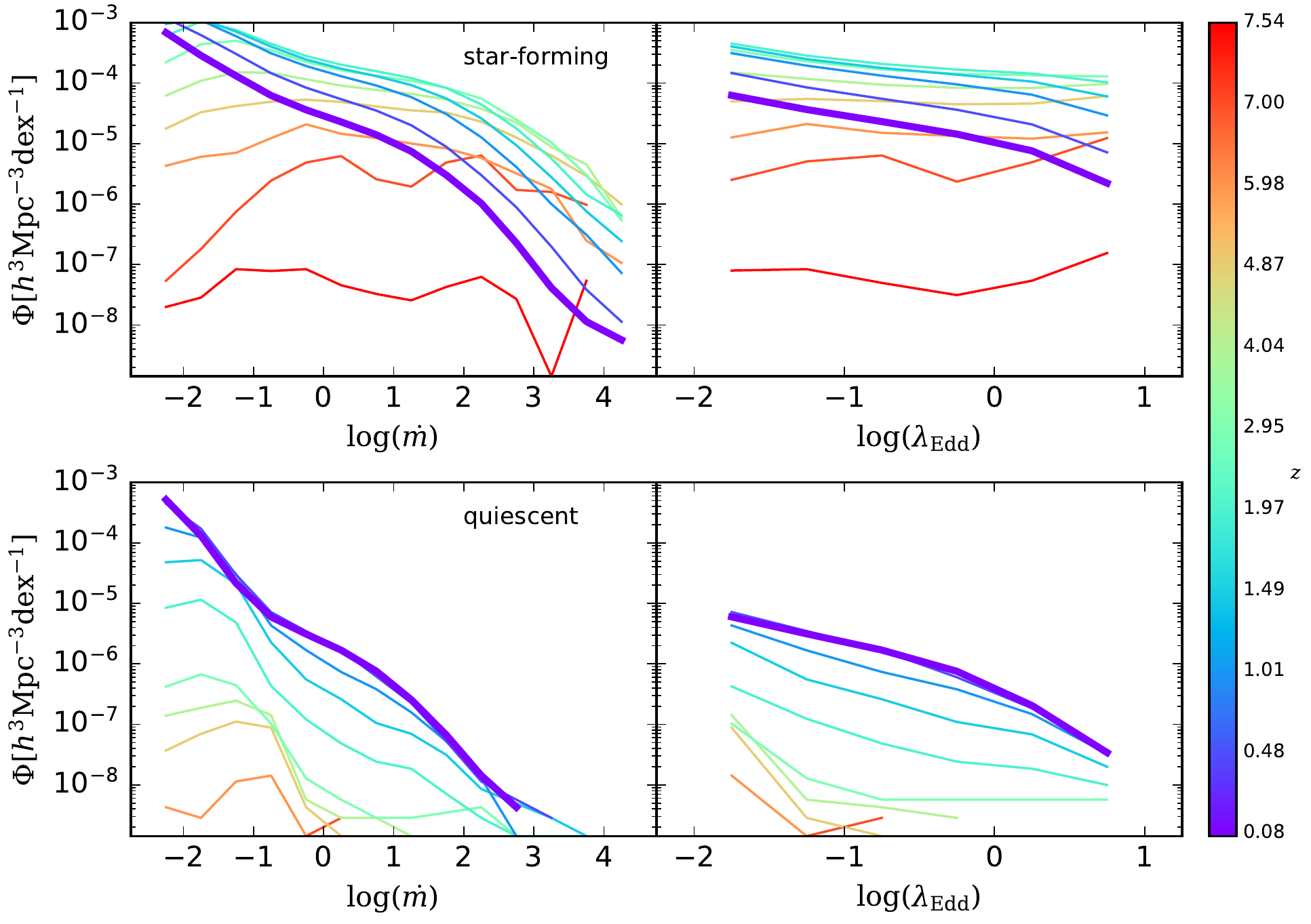}
      \end{center}
			\caption{ERDFs at $0 < z < 8$ of star-forming (sSFR > $10^{-11} \mathrm{yr}^{-1}$) and
			quiescent galaxies (top and bottom panels, respectively.)}
        \label{fig:ERDF_SF}
  \end{figure}

	\cite{Bernhard18}, in contrast to \cite{Bongiorno16},
	suggest the suppression of AGNs with $\lambda_\mathrm{Edd} < 0.1$
	in galaxies with $M_* < 10^{10-11} M_\odot$ to explain both observed AGN LFs and relationship between
	star formation rate and $L_X$.
\cite{Bernhard18} estimate ERDFs separately for star-forming/quiescent host galaxies by
	requiring to explain both observed AGN LFs in hard $X$-ray and stellar mass functions.
	They use the same assumption of $M_\mathrm{BH} = 0.002 M_*$ as \cite{Bongiorno16} for obtaining the Eddington ratio.
	They find that ERDFs of AGNs in star-forming galaxies at $0 < z < 2$ have peaky shape and
	the peak $\lambda_\mathrm{Edd}$ shifts toward higher value with redshift.
	In their estimate, the peak is at $\log(\lambda_\mathrm{Edd}) = -1.7$ at $z \sim 0.2$
	and $\log(\lambda_\mathrm{Edd}) = -0.4$ at $z \sim 2.0$ when the mass-independent ERDFs are assumed.
	However, the Eddington ratio for star-forming host galaxies obtained by our fiducial model
	has a broad distribution (in contrast to the suggestion by \citealt{Bernhard18})
	as shown in Fig. \ref{fig:ERDF_SF}.
	We define star-forming galaxies by their specific star formation rate (sSFR)
	larger than $10^{-11} \mathrm{yr}^{-1}$ \citep{Ilbert13}, which is the similar definition
	to \cite{Bernhard18}.
	The slope of ERDFs for star-forming host galaxies in our model becomes flatter at higher redshift
	in agreement with \cite{Bernhard18} in the sense that
	$\lambda_\mathrm{Edd}$ becomes larger at higher redshift on average.
	Similar to the whole sample (Figs. \ref{fig:ERDF} and \ref{fig:ERDF_bin}),
	the ERDFs of AGNs in star-forming galaxies show that larger $\lambda_\mathrm{Edd}$
	is more common at higher redshift.
	ERDFs for quiescent galaxies, on the other hand, are biased to low Eddington ratio range.
	Also, quiescent galaxies at $z > 5$ do not host AGNs with $L_X > 10^{41}$ erg/s.
	Since the methods for obtaining ERDFs are different, the direct comparison with \cite{Bernhard18}
	is still difficult although it is important for constraining the triggering mechanisms and
	the duty cycle of AGNs.
	We leave further investigation for future studies.

\begin{figure}
			\begin{center}
				\includegraphics[width=\hsize]{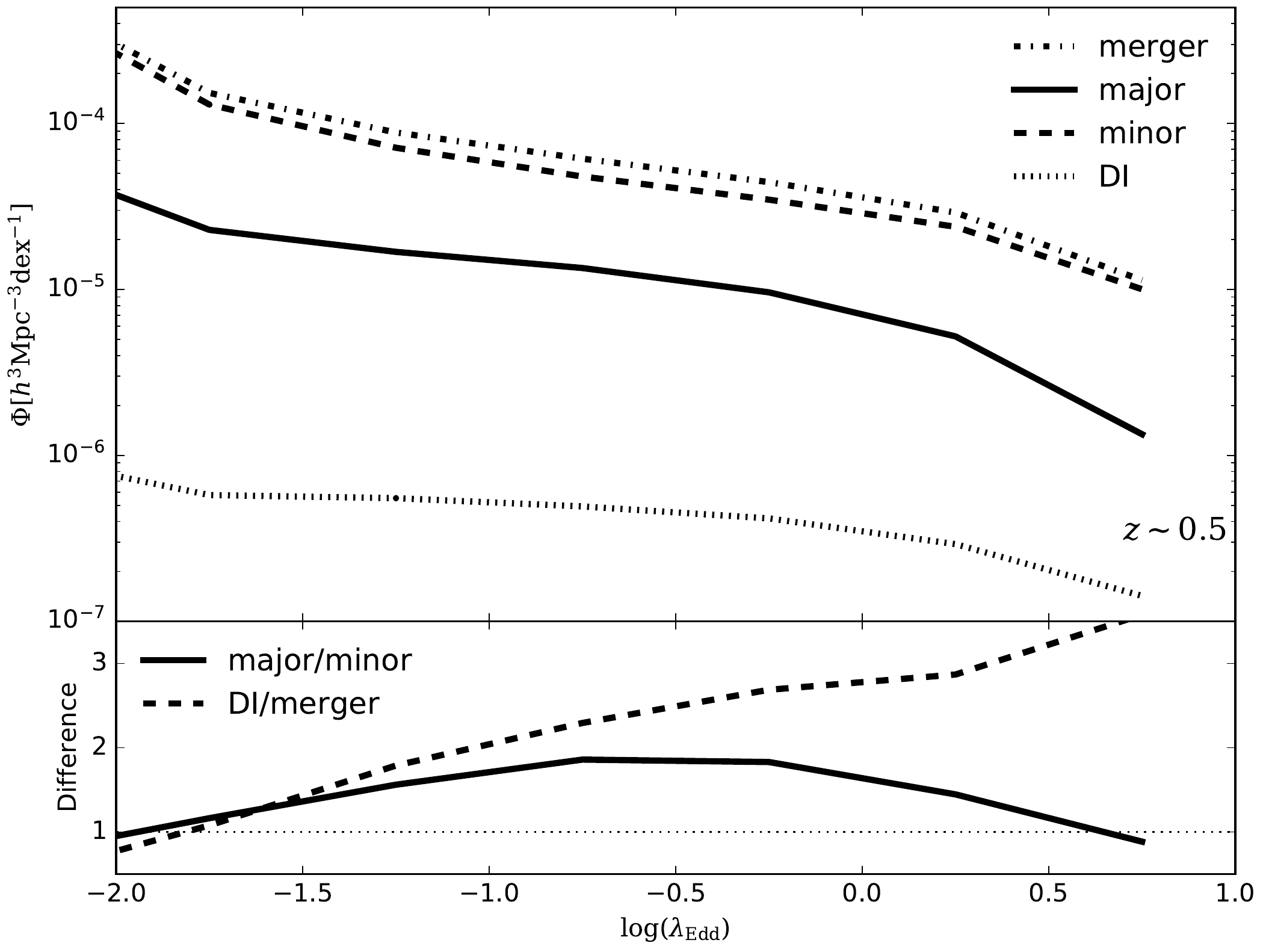}
			\end{center}
			\caption{\textit{Top:} ERDF of AGNs at $z \sim 0.5$, which are triggered by major mergers, minor mergers, mergers (major $+$ minor), and
			disc instabilities (DI).
			These subsamples are described as solid, dashed, dot-dashed, and dotted lines, respectively.
			We define major mergers by the mass ratio of the merging galaxies ($> 0.4$).
			\textit{Bottom:} The difference of the ERDF shape between (1) major and minor mergers (black solid line) and
			(2) mergers and disc instabilities (black dashed line). The $y$ axis shows the ratio of the ERDFs normalised by the
			total number of AGNs which are triggered by each mechanism (i.e. major mergers, minor mergers, mergers, and disc instabilities).
			This panel shows, therefore, only the difference in the shape, not in the normalisation.
			The dotted line means no difference.}
			\label{fig:cum}
	\end{figure}

	Finally, we focus on the difference of the ERDF caused by the different triggering mechanisms of AGNs.
	\cite{Weigel18}, based on a phenomenological model, obtain ERDFs of AGNs triggered by galaxy mergers to reproduce
	observed AGN LFs and the relation between AGN luminosity and the AGN merger fraction.
	They find that although mergers of galaxies are important for triggering AGNs,
	AGNs in merging galaxies do not have higher Eddington ratios compared with those in galaxies without mergers.
	To check the effect of triggering mechanisms on the shape of ERDFs,
	we perform a Kolmogorov-Smirnov test and check the statistical difference of ERDFs in our calculations
	by triggering mechanisms at $z < 4$ (since the sample size is small at larger redshift).
	We compare AGNs with $L_X > 10^{41}$ erg/s and $\log(\lambda_\mathrm{Edd}) > -2.5$
	by their triggering mechanisms in the following two ways:
	(1) major mergers and minor mergers, and
	(2) mergers (major $+$ minor) and disc instabilities.
	We find that in both cases, the $p$-value is less than 3 percent,
	which means that two distributions have different shape at all redshift bins.
	We show the ERDFs of AGNs with different triggering mechanisms at $z \sim 0.5$ in the top panel of Fig. \ref{fig:cum}.
	Dotted line shows AGNs triggered by disc instabilities and dashed and solid lines are AGNs triggered by
	minor and major mergers, and dot dashed line describes the sum of dashed and solid lines (i.e. AGNs triggered by mergers), respectively.
	As for merger-driven AGNs, we find that the ERDF shape is similar independent of the merging mass ratio,
	although the $p$-value is small.
	On the other hand, AGNs triggered by disc instabilities tend to have a flatter distribution.
	To show these differences more clearly, we show the ratio of the normalised ERDFs in the bottom panel of Fig. \ref{fig:cum}.
	Since ERDFs of AGNs triggered by disc instabilities have a flatter shape,
	the difference (DI/merger) becomes larger at larger $\lambda_\mathrm{Edd}$.

	Our result shows that AGNs triggered by major mergers do not have higher $\lambda_\mathrm{Edd}$
	than those triggered by other mechanisms,
	which is broadly consistent with \cite{Weigel18}.
	The relation between the shape of the ERDFs and the triggering mechanisms of AGN activities
	will provide an important information for the cosmic growth of SMBHs and galaxies.
	If the major mergers tend to induce AGNs with higher Eddington ratio,
	it means that the SMBH growth via major mergers are more rapid and the SMBH growth rate will depend on
	the environment of its host galaxy.

	\section{Conclusion}
	\label{sec:conclusion}
	We have presented a theoretical prediction for the ERDFs.
	The calculation is based on our SA model that has
	explained many observational properties of galaxies and AGNs,
	such as LFs, stellar mass functions, $M_\mathrm{BH}$ -- $M_\mathrm{bulge}$ relation,
	and size/velocity -- magnitude relations of galactic discs and bulges.

	We have found that SMBH growths via super-Eddington accretions become more significant
	at higher redshift because the typical gas fraction of the host galaxies is higher at higher redshift.
	This tendency has been also suggested by
	observational studies \citep[e.g.][]{Nobuta12} at $z \lesssim 1.4$.
	Also, \cite{Wu15} show that QSOs at $z > 6$ grow with $\lambda_\mathrm{Edd} \gtrsim 1$.
	Our model provides the detailed evolution of the Eddington ratio distribution
	beyond $z \sim 1.4$ and support the observed trend continues from $z \sim 0$ to $z \sim 8$.
	We have also found that the SMBH growth with the higher Eddington ratio contributes to the SMBH growth
	of less massive SMBHs, indicating that SMBHs
	have grown via super-Eddington accretion.
	We have also found the slowing down of cosmic growth of SMBHs; a trend that
	higher $\lambda_\mathrm{Edd}$ ranges present their peak number density at higher redshift.

	To compare the model results with observations and to understand the SMBH growth,
	it is important to take the observed sample selection into account.
	Shallower observations tend to underestimate the number density of AGNs
	not only with lower but also higher Eddington ratios.
	The slope of ERDFs, therefore, is sensitive to the sample selection.
	ERDFs of AGNs with different properties of their host galaxies
	would provide the information on the co-evolution.
	Such detailed analysis will be observationally available in the
	era of deep spectroscopic surveys with a larger sample size.

  \section*{Acknowledgements}
		We appreciate the detailed review and useful suggestions
		by the referee, which have improved our paper.
		We appreciate the fruitful comments from the observational side
		by M. Akiyama, W. He, and T. Kawamuro.
		We also thank T. Ishiyama, M. A. R. Kobayashi, R. Makiya, M. Enoki,
		and K. Okoshi for the development of \nugc~and discussions.
		H. ~Shirakata has been supported by the Sasakawa Scientific Research Grant from
		the Japan Science Society (29-214), JSPS KAKENHI (18J12081), and
		a grant from the Hayakawa Satio Fund awarded by the Astronomical Society of Japan.
		T. ~Kawaguchi was supported in part by JSPS KAKENHI (17K05389),
		grant from the Japan research institute of industrial science,
		and Japan Foundation for Promotion of Astronomy.
		T.~Okamoto is financially supported by
		MEXT KAKENHI Grant (18H04333).
		M.~Nagashima has been supported by the Grant-in-Aid (25287041 and 17H02867)
		from the MEXT of Japan.
		This work was also supported in part by World Premier International Research Center Initiative (WPI),
		MEXT, Japan, by MEXT Priority Issue 9 on Post-K Computer
		(Elucidation of the Fundamental Laws and Evolution of the Universe), and by JICFuS.




  \bibliographystyle{mnras}
  \bibliography{Astrophysics} 

  \bsp	
  \label{lastpage}
\end{document}